\documentclass[aps,prb,preprint,groupedaddress]{revtex4-1}
\usepackage{color}
\usepackage{graphicx,amsmath}
\usepackage{epstopdf}

\usepackage[normalem]{ulem}

\begin{document}

\title{Twisted magnetic patterns: \\
Exploring the Dzyaloshinskii--Moriya vector}

\author{V. E. Dmitrienko}
\affiliation{Institute of Crystallography, Moscow, Russia}
\author{E. N. Ovchinnikova}
\affiliation{Physical Department,
Moscow State University, Russia}
\author{S. P. Collins}
\affiliation{Diamond Light Source Ltd, Diamond House, Harwell Science and Innovation Campus, Didcot, Oxfordshire, OX11 0DE, UK}
\author{G. Nisbet}
\affiliation{Diamond Light Source Ltd, Diamond House, Harwell Science and Innovation Campus, Didcot, Oxfordshire, OX11 0DE, UK}
\author{G. Beutier}
\affiliation{SIMaP, UMR 5266 CNRS, Grenoble-INP UJF, France}
\author{Y.O. Kvashnin}
\affiliation{European Synchrotron Radiation Facility, 6 Rue Jules Horowitz, BP220, 38043 Grenoble Cedex, France}
\author{V. V. Mazurenko}
\affiliation{Department of Theoretical Physics and Applied Mathematics, Ural Federal University,
Mira str. 19, 620002 Ekaterinburg, Russia}
\author{A.I. Lichtenstein}
\affiliation{I. Institut f{\"u}r Theoretische Physik, Universit{\"a}t
Hamburg, Jungiusstra{\ss}e 9, D-20355 Hamburg, Germany}
\author{M.I. Katsnelson}
\affiliation{Radboud University Nijmegen, Institute for Molecules and Materials, Heyendaalseweg 135, NL-6525 AJ Nijmegen, The Netherlands}
\date{\today}
\maketitle

{\bf 
Magnetism - the spontaneous alignment of atomic moments in a material - is driven by quantum-mechanical `exchange' interactions which 
operate over atomic distances as a result of the fundamental symmetry of electrons. Currently, one of the most active fields of 
condensed matter physics involves the study of magnetic interactions that cause 
\cite{Eerenstein06,Cheong07}
, or are caused by 
\cite{Rosler06,Yu10}
a twisting of nearby atoms.
This can lead to the magnetoelectric effect that couples electric and magnetic properties, and is predicted to play a prominent role in future 
technology. 
Here, we discuss the complex relativistic interplay between magnetism and atomic crystal structure in a class of materials called `weak ferromagnets'. 
The sign of the underpinning \emph{Dzyaloshinskii--Moriya}\cite{Dzyaloshinskii57, Dzyaloshinskii58, Moriya60a, Moriya60b} interaction has been determined for the first time, by using synchrotron 
radiation to study iron borate (FeBO$_3$). We present a novel experimental technique based on interference between two x-ray scattering processes 
(one acts as a reference wave) which we combine with a second unusual approach of turning the atomic antiferromagnetic motif with a small magnetic field.
We show that the experimental results provide a clear validation of state-of-the-art theoretical calculations. 
These experimental and theoretical approaches open up new possibilities for exploring, modelling and exploiting novel magnetic and 
magnetoelectric materials.
}

There is considerable mystery behind the origins of complicated structures. 
While the dominant short-range interactions that allow 
the building blocks to grow are well understood, the much more subtle forces that lead to a particular twisting at larger length-scales, 
such
as chiral biological molecules and liquid crystals \cite{Kitzerow01}, and canted magnetic systems \cite{Rosler06}, remain subjects of topical debate. 
In this Letter we seek to address this question for the case of magnetism.
Our main findings are two-fold: first, we demonstrate a novel and elegant experimental method for exploring magnetic 
materials with weak relativistic spin-orbit interactions, and second, we present a state-of-the-art quantum-mechanical many-body approach to 
the detailed description of such interactions in crystals. As a touchstone example we selected the crystal of iron borate (FeBO$_3$) 
which is a strongly correlated electron system with a relatively simple crystal structure, nonetheless allowing a non-trivial canted 
and locally twisted magnetic ordering pattern. Taken together, these two strands demonstrate that modern condensed matter theory is 
capable of determining the elusive sense, or sign, of the \emph{Dzyaloshinskii-Moriya} (DM) interaction, and is thus able to elucidate 
the mechanism for coupling electric and magnetic degrees of freedom in magnetoelectric multiferroics, and to begin to predict the 
properties of this important class of materials.

The interactions between atomic magnetic moments (or spins) is not direct, but mediated by the intervening matter. 
Coupling can be diminished through screening\cite{Blundell01}, or enhanced, for example, by superexchange 
via oxygen atoms\cite{White07}. Moreover, the coupling is a property of the material and must therefore possess all of its symmetries (Neumann's principle).
The most general form of the bi-linear coupling energy between two spins 
contains a scalar (isotropic) exchange term, exchange anisotropy (which we will neglect for the present discussion)
and an antisymmetric term that reverses with permutation of the spin indices.
The latter is the Dzyaloshinskii-Moriya interaction, 
which can be expressed in terms of a DM vector, $\mathbf{D}$, and the vector product of spins, leading to a spin-dependent energy (classical Heisenberg model),
\begin{equation} \label{eq:H-DM}
\Delta E= \sum_{m\neq n} J_{mn} \mathbf{S}_m\cdot \mathbf{S}_n +  \sum_{m\neq n} \mathbf{D}_{mn}\cdot [ \mathbf{S}_m \times \mathbf{S}_n]  - \mu_BgS\sum_{m}\mathbf{H}\cdot \mathbf{S}_m
\end{equation}
where the summations run over all magnetic atoms. $\mathbf{S}_m$ is a unit vector along the direction of the $m^{th}$ spin of magnitude $S$,
$\mu_B$ is the Bohr magneton 
and $g\approx -2$ is the gyromagnetic ratio. Both the exchange coefficients $J_{mn}$ and the DM vectors $\mathbf{D}_{mn}$ depend 
on the relative positions of the magnetic atoms. The first term in equation (\ref{eq:H-DM}) prefers parallel/antiparallel coupling of spins (depending on the 
sign of $J_{mn}$), while second favours an orthogonal alignment, producing a twist, or canting, of the atomic moments.  

In $3d$ transition metal oxides it is usually necessary to consider only the nearest neighbours, since these dominate the exchange 
interactions. The DM interaction is typically a few percent of the isotropic term ($\sim$0.1 meV, compared to $\sim$10 meV), producing 
just a modest canting. Nevertheless, the effect is important. A spontaneous rearrangement of atoms to favour the DM 
interaction (often called the inverse DM
effect) can produce a large electric polarization in magnetoelectric materials\cite{Eerenstein06,Cheong07}. In so-called `weak ferromagnets' canting of the otherwise 
collinear antiferromagnetic arrangement leads to a small net ferromagnetic polarization that couples strongly to an external magnetic field\cite{Matarrese54}.

Important symmetry restrictions on the DM vector have been discussed since it was initially  introduced  more than fifty years ago
\cite{Dzyaloshinskii57,Dzyaloshinskii58, Moriya60a,Moriya60b}.
In the case of FeBO$_3$, with crystal symmetry $R\bar{3}c$, there are two iron atoms at the $2b$ inversion centres,
two boron atoms at the $2a$ positions, and six oxygens at $6e$ sites.
If the oxygen atoms were absent, or positioned exactly between two neighbouring iron atoms, the symmetry would be $R\bar{3}m$, which does not allow the
DM interaction. It is therefore the small displacement of oxygen atoms, striving towards a close-packed structure, that drives iron borate into the observed 
complex magnetic ordering pattern.
A closer examination of the FeBO$_3$ crystal structure 
(Fig. 1)
reveals that each Fe atom is connected to six equivalent nearest Fe neighbours: three in the plane
above and three below. The six DM vectors linking these Fe atoms are related by symmetry and when summed, lead to a resultant vector along $z$.
It follows from equation (\ref{eq:H-DM}) that the DM 
vector of this type induces a twist between $A$ and $B$ spins in the $xy$ plane, but the symmetry alone cannot say whether 
this twist will be left-handed or right-handed. The absolute sign of local twist can be found both experimentally and theoretically 
using techniques described in the following text. 

The present experimental technique relies on the fact that the weak ferromagnetic moment is perpendicular to the opposing antiferromagnetic (AF)
components 
(Fig. 1)
. Since the former can be turned with a small external field (Fig. 2), the dominant AF structure is `dragged' around to follow it, offering a
powerful new method to manipulate the magnetic x-ray diffraction.
In order to determine the sign of the DM interaction, we must determine whether the rotation of spins is in the same sense as the rotation of 
the oxygen triangles, or opposite. Unfortunately, the standard techniques for characterising antiferromagnetic structures - x-ray or 
neutron diffraction - do not help: The sign of the twist appears in the $phase$ of the diffracted wave, which is lost in an intensity 
measurement - an aspect of the famous `phase problem' of crystallography. Borrowing from the ideas behind holography, it was recently
suggested by some of us\cite{Dmitrienko2010} that the sign of the DM vector could be measured with resonant x-ray diffraction by observing 
interference between the resonant\cite{Blume94} and magnetic\cite{deBergevin1981} scattering amplitudes.
The resonant scattering process adopted is a rather exotic one, involving pure electric quadrupole events ({\it i.e.} beyond the usual 
dipole approximation). Its phase and amplitude vary rapidly with photon energy, being significant only very close the the Fe $K$ x-ray absorption edge 
energy of 7.11 keV, and it has a complex dependence on both photon polarization and the rotation of the sample about the normal to the 
diffracting planes ($\psi$-angle). However, in recent years, such phenomena have been studied in detail and are now extremely well 
understood \cite{Dmitrienko1,Lovesey1}. Moreover, both the resonant and magnetic scattering signals appear at the same Bragg reflection positions - 
$(hkl)=(0,0,6n+3)$ - that are `forbidden' for 
the vastly stronger charge scattering processes, and have comparable amplitudes to each other, maximizing the effects of interference. 
The sign and amplitude of the magnetic scattering signal depends on the spin direction, which can be rotated with a magnetic field. 
We thus expect control of the amplitude and phase of both the magnetic scattering 
and resonant reference wave, 
allowing the phase of the magnetic scattering to be determined.
Details of the magnetic and resonant scattering 
amplitudes are given in SI.

Three types of measurement are presented. The first shows a remarkable effect - an apparent jump in the energy of the resonant scattering
peak as the magnetic motif is rotated by 180$^\circ$, as a result of constructive (destructive) interference 
to the low (high) energy side of the resonance 
(Fig. 3)
. The opposite jump was observed when the phase of the resonant
scattering was reversed by changing the sample $\psi$ angle. Both jumps are reproduced by our {\it ab initio} 
calculations, which make a definite prediction for the sign of the DM interaction, the phase of the magnetic scattering amplitude, 
and thus the direction of the jump. The second measurement shows the intensity, measured as a continuous rotation of the field angle, 
for the low and high energy side of the resonance. For the final measurement, the sample azimuthal angle was varied continuously, 
with a fixed photon energy and field applied in two orthogonal directions. In all cases, the phase of the magnetic scattering, 
who's reversal would convert red to green lines and {\it vice versa} were consistent, completely unambiguous, and in agreement 
with the calculations. 

One of the main goals of the present work is to demonstrate that the sign of the DM interaction can be determined reliably 
not only by experiment but also theoretically.
To this end we have performed first-principles calculations by using Local Density 
Approximation incorporating the on-site Coulomb interaction U and the Spin-Orbit coupling (LDA+U+SO)
\cite{shorikov-lda-u-so,solovyev-lda-u-so}. 
Our calculations predict (see SI) that the lowest energy stable magnetic structure is precisely the one observed experimentally 
and shown in 
Fig. 1.
Furthermore, we predict that the magnetic twist between adjacent layers is in the {\it same} direction as the twist of the 
oxygen triangles. This is the basis for the calculated
curves in 
Fig. 3
, and is thus very clearly confirmed by experiment. 

Since the vector product 
$[\mathbf{S}_A \times \mathbf{S}_B]$
in equation~(\ref{eq:H-DM}) is parallel to z-axis, and the corresponding 
DM interaction 
must {\it reduce} the energy of the system,
we can deduce that 
$D^{z}_{AB}$ is {\it negative}. The absolute value of the DM interaction energy is readily estimated from the measured canting 
angle (0.9$^{\circ}$) and isotropic exchange interaction: $|D^{z}_{AB}| = 2 J| \frac{S_y}{S_x}| $ =   0.33 meV. Here $J$
is approximately 10.3 meV.

The determination of the $J_{mn}$ and $\vec D_{mn}$ parameters 
with account of hybridization, correlation, temperature and spin-orbit coupling effects is a complex methodological and computational 
problem
requiring
a whole arsenal of numerical techniques \cite{Moriya60b, solovyev-lamno3, aharony, mazurenko-fe2o3}.
Here, we outline a second, and very general next-generation
method that has a simple formulation and captures all important electronic and magnetic excitation effects 
\cite{katsnelson}. The resulting expression for the {\it correlated band} DM interaction can 
be applied to a wide range of materials 
and is given by 
\begin{eqnarray}
\mathbf{D}_{mn} = -\frac{i}{2} Tr_{L,\sigma}\biggl\{N_{nm} [\hat{\mathbf{J}}, \hat t_{mn}]_{+}\biggl\}, \label{vecD}
\end{eqnarray}
where $N_{nm}$ is the the energy-integrated inter-site Green's function which describes the propagation of an 
electron from site $n$ to $m$, $\hat{\mathbf{ J}}$ is the total moment operator, $t_{mn}$ is the hopping  matrix and $[.,.]_{+}$ represents an anticommutator (see SI 
\S IV for details).
The method is developed in a many-body form, thus, the state-of-the-art numerical approaches such as dynamical mean-field theory can 
be used to take into account temperature and  dynamical Coulomb correlations effects. 

The calculated DM vector linking iron atoms 0 and 1 
(Fig. 1)
, for example, is $D_{01}$=(-0.249, 0, -0.240) meV. By symmetry, 
it lies in the $xz$ plane, perpendicular to the two-fold axis that passes through the oxygen atom\cite{Moriya60b}. All six 
symmetrically-equivalent vectors have the same $z$ component, but the $xy$ components average to zero.
Our calculated canting angle of 0.7$^{\circ}$ is only slightly smaller than
experimental value of 0.9$^{\circ}$, used for the self-consistent calculation. 

Crucially, the sign of the DM interaction, which we have predicted by two theoretical methods, determines the direction of twist of the magnetic
structure, which affects the phase of the magnetic scattering and the sign of the interference term in 
Fig. 3.
It is thus confirmed unambiguously by the experimental data.

In conclusion, we show that 
a new interference technique in which
measurements are carried out with precise control of the amplitude and phase of a reference wave
gives an unambiguous result for the sign of the Dzyaloshinskii-Moriya interaction. We find that 
the twist in the magnetic structure is in the {\it same} direction as that of the oxygen atoms.
The results prove the efficacy 
of state-of-the-art electronic structure calculations, able to predict the magnetic ground-state and both direction and strength of the 
DM interaction.
These findings take us a step closer to realising the prediction of complex non-collinear magnetic structures and the associated properties of
an important class of materials that includes weak ferromagnets and multiferroics.


{\bf Experimental Method}

The experiments were mostly carried out at beamline BM28 (XMaS), European Synchrotron Radiation Facility \cite{BM28}, with preliminary investigations 
of the pure magnetic and pure resonant scattering carried out at beamline I16, Diamond Light Source \cite{I16} (See SI). 
Both beamlines provide intense x-ray beams covering the required energy range ($\approx 7$ keV), focussed  onto a 10-800K cryofurnace, 
mounted at the centre of large six-circle diffractometers. The sample - a single crystal of FeBO$_3$ $\sim 4 \times 3 \times .05$ mm$^3$ in size - was attached 
with its 001 surface normal to the diffractometer $\phi$ rotation axis \cite{BM28}, which was parallel to the azimuthal rotation axis, $\psi$ 
(Fig. 2).
Two small rare-earth magnets provided a magnetic field of 0.011 T, sufficient to 
saturate the weak ferromagnetism within the crystal $xy$ plane, with an orientation determined by the motorized rotation angle of the magnets around the 
$\phi$ axis. Scattering was in the vertical plane, perpendicular to the linearly ($\sigma$)-polarized incident beam, 
and a linear polarization analyser, based on a Cu 220 Bragg 
reflection, selected just the rotated ($\sigma \rightarrow \pi$) polarization channel. Most measurements were carried out 
at temperature T=200K where the moments are close to saturation, with subsidiary measurements performed at T=400K (well above the magnetic ordering temperature of 
$\sim$ 348 K)
where the magnetic scattering is absent.

\newpage
{\bf Figure Legends}

\begin{figure}[!h] 
\begin{center}
\includegraphics[angle=270,width=80mm]{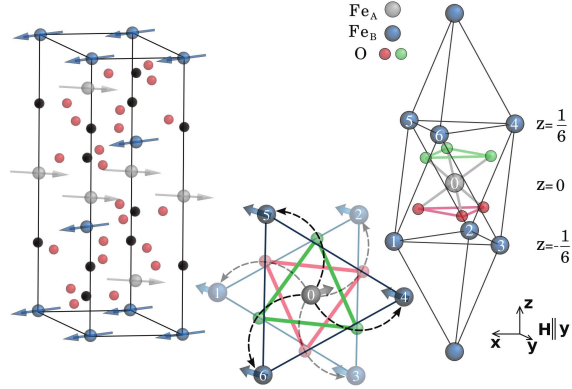}
\caption{\small
Atomic and magnetic order in FeBO$_3$. 
Left: a magnetic (hexagonal) unit cell, showing oxygen atoms (red), boron atoms (black), and two symmetry-related magnetic iron
sublattices (blue and grey) with moments tilted between the two.
Right: The local environment of one of the grey (A-site) Fe atoms, showing neighbouring B-site Fe atoms (blue). The upper and lower oxygen triangles are
coloured green and red, and boron atoms removed, for clarity.
Centre: The same structure from the top, highlighting the twisted superexchange paths from the A-site Fe atom to to upper Fe layer (dark blue)
and the lower layer (pale blue) via the oxygen triangles.
}
\label{febo3_atoms}
\end{center}
\end{figure}

\begin{figure}[!h] 
\begin{center}
\includegraphics[angle=270,width=50mm]{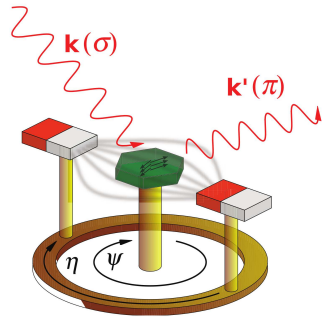}
\caption{\small
A schematic of the experiment, showing the incident and scattered x-ray beams and the FeBO$_3$ crystal with weak ferromagnetic moments
directed along the direction of the applied magnetic field. Sample and field were rotated about a common axis so that 
$\psi=0$ and $\eta$=0 correspond to the the crystallographic (100)$_{hex}$ axis, and 
field direction (from south to north pole of the magnet assembly), directed along ${\mathbf k}+{\mathbf k}^{\prime}$, respectively.
 }
\label{febo3_experiment}
\end{center}
\end{figure}

 \begin{figure}[!] 
\begin{center}
\includegraphics[angle=0,width=90mm]{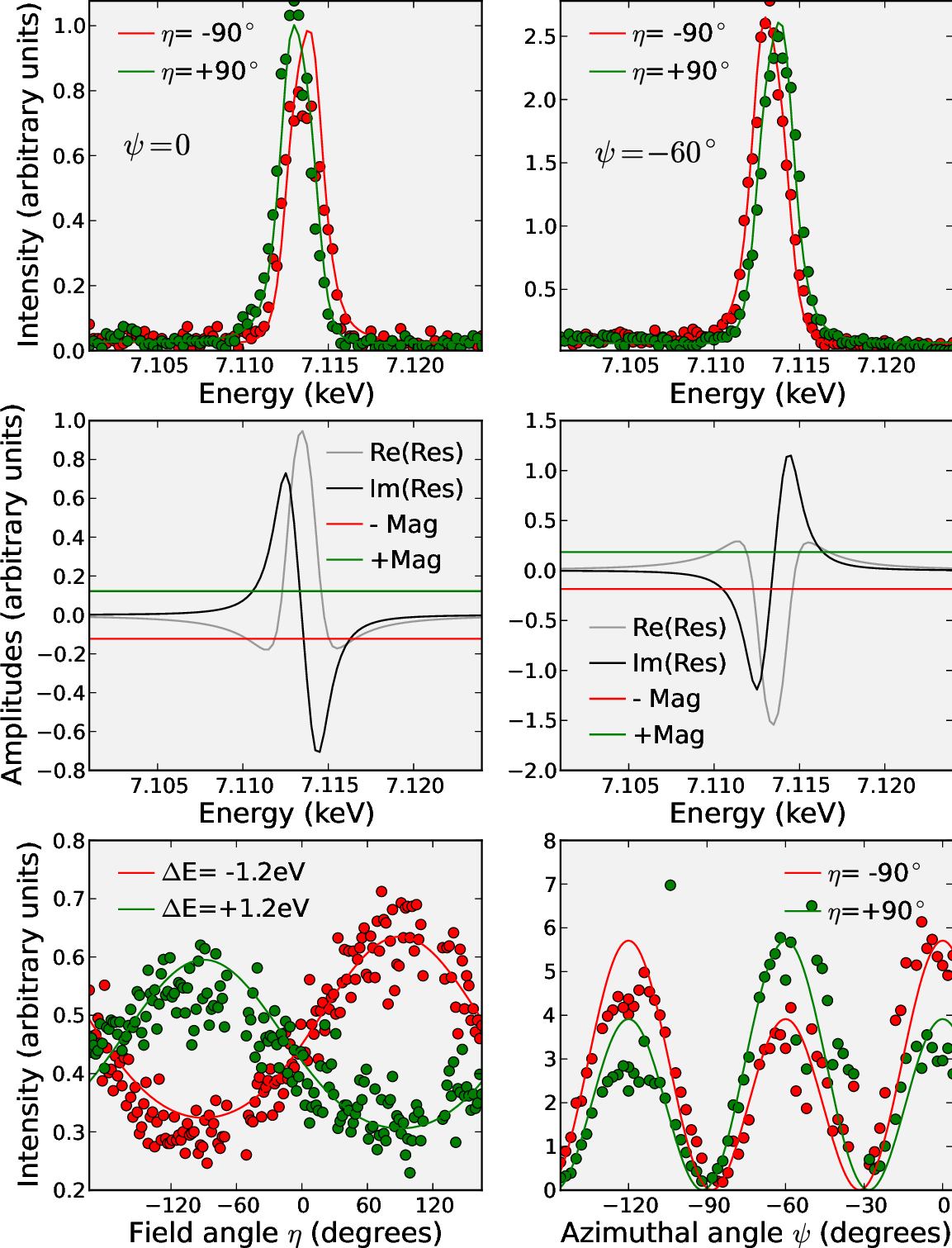}
\caption{\small
The major experimental results from FeBO$_3$ along with simulations based on a `double resonance' model (see S.I).
Top: A remarkable shift in the x-ray resonance energy is observed on rotation of the magnet ($\eta=\pm90^{\circ}$). The shift reverses
as the sample azimuth is rotated from $\psi=0$ to $\psi=60^{\circ}$, and is evident in both experimental data and model calculations.
Middle: The origin of the energy jump can be understood by considering interference between the constituent amplitudes. The magnetic amplitude
is in phase with the {\it imaginary} part of the resonance amplitude (black lines), causing constructive/destructive
interference on the low/high energy side of the resonance, depending on the relative phase of the magnetic and resonant scattering amplitudes.
Bottom: The same fitted amplitudes give very good agreement with intensity measurements {\it versus} field angle (left), with the resonance amplitude reversed by
shifting the energy above and below the resonant centre. Right: The intensity variation with azimuthal angle and opposite magnetic amplitudes ($\eta=\pm90^{\circ}$).
In all plots, circles represent experimental data and solid lines are calculations.
}
\label{febo3_results}
\end{center}
\end{figure}

{\bf Acknowledgments}

The work of VVM is supported by the grant program of President of Russian Federation
MK-5565.2013.2, the contracts of the Ministry of education and science of Russia N 14.A18.21.0076 and 14.A18.21.0889.
VED is grateful for the grant of Presidium RAS ``Diffraction of synchrotron radiation in multiferroics and chiral magnetics''.
MIK acknowledges a financial support by FOM (The Netherlands). We thank the staff of ESRF BM-28 for expert assistance, and 
Yuri Shvyd'ko for the loan of the FeBO$_3$ crystal.

\newpage
\begin{center}
\large
{\bf Twisted magnetic patterns: \\
Exploring the Dzyaloshinskii--Moriya vector\\
Supplementary Information}
\end{center}

\normalsize
\section{Pure Magnetic Scattering from Iron Borate}
\label{puremag}

The primary result of the present study is to determine the phase of magnetic x-ray scattering from FeBO$_3$, by observing interference with resonant forbidden
scattering, from which the sign of the Dzyaloshinskii-Moriya (DM) interaction can be determined and compared to new theoretical models. In order to carry out such an analysis with
confidence, it is necessary to establish that the pure magnetic and pure resonant signals can be modelled reliably and accurately, 
with a particular emphasis on their phases. We begin by discussing the theoretical and experimental forms of the magnetic scattering.

Quantum or semi-classical electrodynamical calculations give a well established expression for the
amplitude of {\it spin} magnetic x-ray scattering 
\cite{SIdeBergevin1981,SIBlume1985,SILovesey1996} 
\begin{equation} 
f_\mathrm{m}= ir_e\frac{\hbar\omega}{mc^2}\mathbf{B}\cdot\mathbf{F}_\mathrm{spin}(\mathbf{Q})
\label{eq:fm}
\end{equation}
where $\mathbf{Q}$ is the scattering vector, vector $\mathbf{F}_\mathrm{spin}(\mathbf{Q})$ is the corresponding Fourier harmonic of the spin (not magnetic moment!) 
density, $r_e=e^2/mc^2$ is the classical electron radius,  $\hbar\omega$ is the photon energy and vector $\mathbf{B}$ determines polarization properties of spin 
scattering: \begin{equation} \label{eq:B}
\mathbf{B}=(\boldsymbol{\epsilon'^{*}}\times\boldsymbol{\epsilon}) - \left[(\mathbf{k'}\times\boldsymbol{\epsilon'^{*}})\times(\mathbf{k}\times\boldsymbol{\epsilon})
- (\mathbf{k'}\times\boldsymbol{\epsilon'^{*}})(\mathbf{k'}\cdot\boldsymbol{\epsilon})+(\mathbf{k}\cdot\boldsymbol{\epsilon'^{*}})(\mathbf{k}\times\boldsymbol{\epsilon})
\right]/k^2.
\end{equation}
Here $\boldsymbol{\epsilon}$ ($\boldsymbol{\epsilon'}$) and $\mathbf{k}$ ($\mathbf{k'}$) are the polarization and wave vectors of the incident (scattered) waves, 
$\mathbf{k'}=\mathbf{k}+\mathbf{Q}$, $k=\omega/c$. 
The sign of this 
expression is positive if the x-ray plane wave is written as $\exp[i(\mathbf{k}\cdot \mathbf{r}-\omega t)]$. The (negative) sign of the scattering electronic charge, $e$, 
is not important because the scattering amplitude depends only on $e^2$. For $\boldsymbol{\sigma}$ and $\boldsymbol{\pi}$ polarizations (linear polarization 
perpendicular and parallel to the scattering plane, $\boldsymbol{\sigma}\times\boldsymbol{\pi}=\mathbf{k}$) and polarization states expressed as
column vectors, $\mathbf{B}$ can be written as
\begin{equation} \label{eq:B}
\mathbf{B}(\boldsymbol{\epsilon'},\boldsymbol{\epsilon})
=\left(\begin{array}{cc}
\mathbf{B}_{\sigma'\sigma} & \mathbf{B}_{\sigma'\pi} \\
\mathbf{B}_{\pi'\sigma} & \mathbf{B}_{\pi'\pi}
\end{array}\right)=\left(\begin{array}{cc}
\mathbf{k}\times\mathbf{k'} & -\mathbf{k'}(1-\mathbf{k}\cdot\mathbf{k'}) \\
\mathbf{k}(1-\mathbf{k}\cdot\mathbf{k'}) & \mathbf{k}\times\mathbf{k'}
\end{array}\right).
\end{equation}
It is important to note that the non-resonant magnetic scattering  has both a non-rotated 
($\boldsymbol{\sigma} \rightarrow\boldsymbol{\sigma}^{\prime}$) and a rotated ($\boldsymbol{\sigma} \rightarrow\boldsymbol{\pi}^{\prime}$) component. Only the latter will play a 
role in interference because the resonant scattering has only this component. The orbital contribution to non-resonant magnetic scattering is expected to be small for 
iron in FeBO$_3$ (confirmed both by our experimental data and {\it ab initio} simulations) and will be neglected for the time being.

In weak ferromagnets of FeBO$_3$ type, the direction of $\mathbf{F}_\mathrm{spin}(\mathbf{Q})$ for the pure antiferromagnetic reflections is normal to the threefold axis and to the direction of the 
external magnetic field $\mathbf{H}$ (applied in the easy plane). There are only two iron atoms per unit cell and the magnetic structure factor 
$\mathbf{F}_\mathrm{spin}(\mathbf{Q})$ of forbidden reflections ({\it i.e.} reflections of the form $(hkl)=(0,0,6n+3)$, which are completely forbidden by 
spacegroup selection rules for isotropic scattering) has a simple form: 
$\mathbf{F}_\mathrm{spin}(\mathbf{Q})=2Sf_S(\mathbf{Q}) \frac{\mathbf{H}\times\mathbf{\bar{D}}}{|\mathbf{H}||\mathbf{\bar{D}}|}$ 
where $S$ is the total spin of iron atom and $f_S(\mathbf{Q})$ is the spin form factor. If the spin structure factor 
$\mathbf{F}_\mathrm{spin}(\mathbf{Q})$ is real (for instance for centrosymmetric structures) then the magnetic 
scattering amplitude is purely imaginary. {\it The sense (sign) of  $\mathbf{F}_\mathrm{spin}(\mathbf{Q})$ is determined by the sign of the 
average Dzyaloshinskii-Moriya vector $\mathbf{\bar{D}}$ and by direction of the magnetic field $\mathbf{H}$}. Vector 
$\mathbf{F}_\mathrm{spin}(
\mathbf{Q})$ can be rotated in the easy plane by rotation of $\mathbf{H}$ as long as $\mathbf{H}$ is strong enough to overcome 
the in-plane magnetocrystalline anisotropy (in the $xy$ plane).

\begin{figure}[!h] 
\begin{center}
\includegraphics[angle=0,width=80mm]{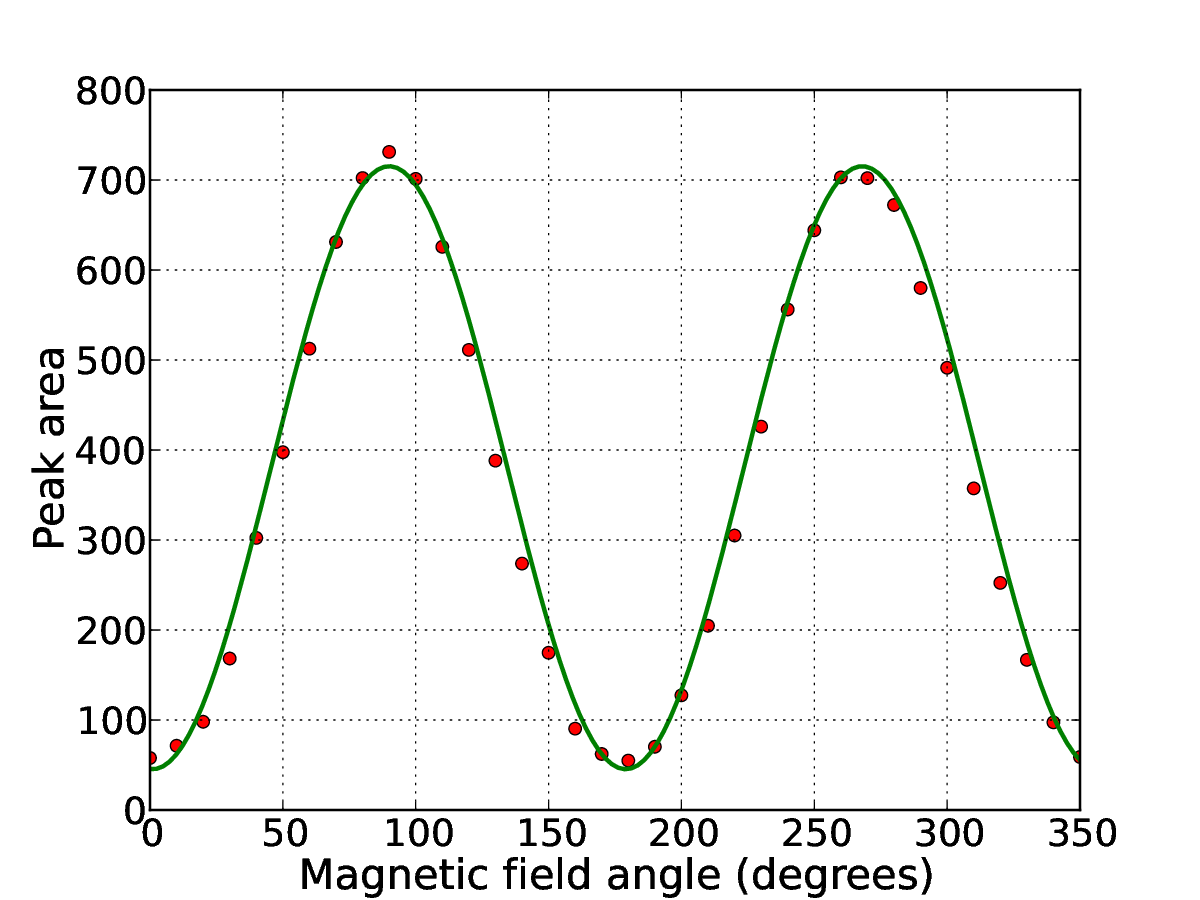}
\caption{\small
The dependence of the non-resonant magnetic (003) Bragg reflection intensity on the magnetic field angle (B$\simeq$0.011 T)), measured at an energy of 5.1 keV. Red circles are data points and the calculated
spin-scattering intensity is shown by a solid green line.
}
\label{FeBO3_vs_fieldangle_new}
\end{center}
\end{figure}

\begin{figure}[!h] 
\begin{center}
\includegraphics[angle=0,width=80mm]{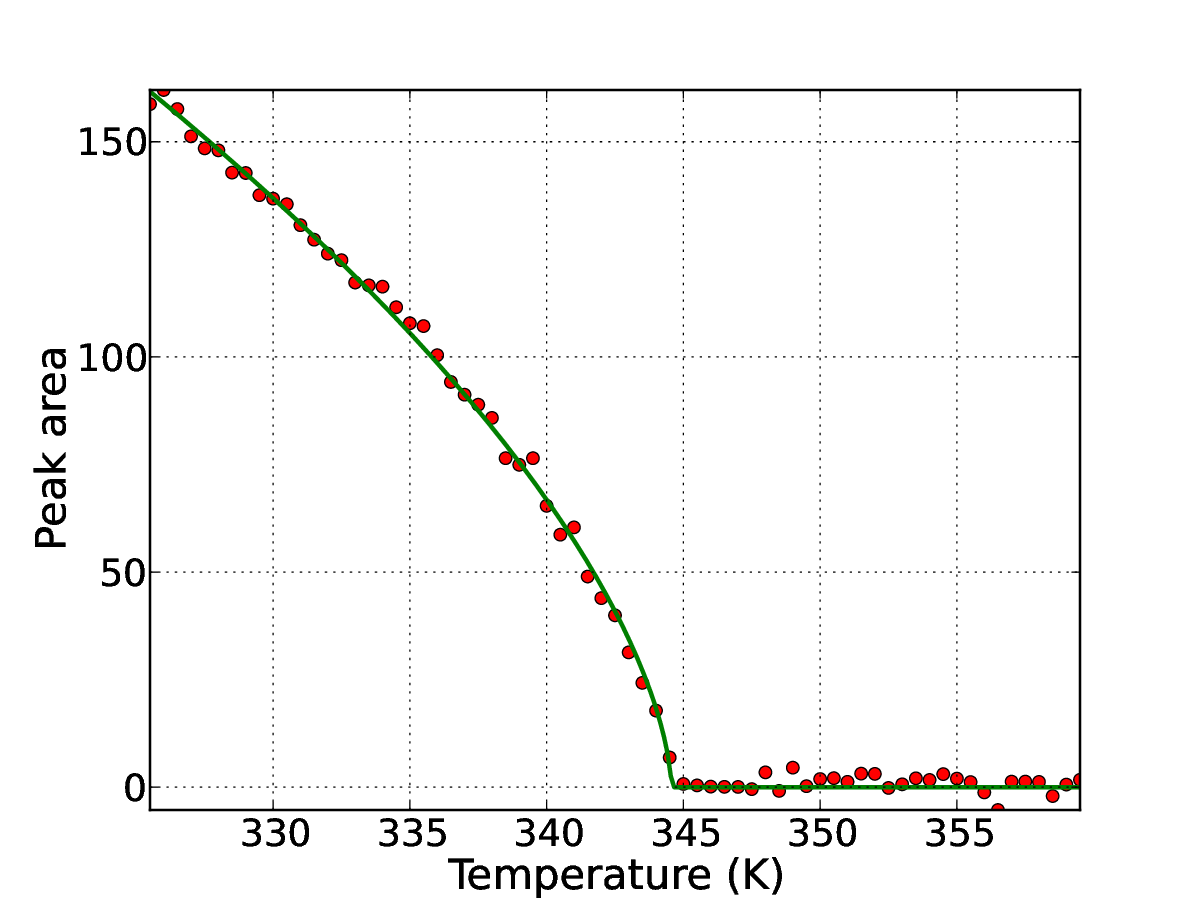}
\caption{\small
The (003) magnetic diffraction intensity at 5.1 keV with a field angle of $90^\circ$ {\it vs} temperature, across the magnetic phase transition. The solid line is a power-law fit to the function $I\propto |M|^2 \propto (T_N-T)^{2\beta}$, giving $\beta=0.305(2)$.
}
\label{FeBO3_vs_temp}
\end{center}
\end{figure}

Measurements of the pure magnetic (003) reflection {\it vs} magnetic field angle were carried out at Beamline I16, Diamond Light Source \cite{SII16}, 
at an energy of
5.1 keV, chosen to be far from the iron $K$-edge resonance and free from multiple scattering artefacts. The results, shown in 
Fig. \ref{FeBO3_vs_fieldangle_new}, reveal data of spectacularly
high quality due to that fact that no sample rotation is required for this new type of measurement. Moreover, the data agree extremely well with the calculated intensity
(to within a single scaling parameter) based on the above expressions, confirming the negligible contribution from orbital magnetism. 
Magnetic scattering measurements were carried out at ambient temperature, well below the ordering temperature of T$_N\simeq$ 348 K. On heating 
the sample in a closed-cycle cryofurnace, the magnetic
scattering intensity followed the expected form of a second-order phase transition (Fig. \ref{FeBO3_vs_temp}), reported in the literature \cite{SIEicschutz1970}.

\section{Pure Quadrupole Resonant Scattering from Iron Borate}

The charge scattering amplitude consists of two parts: non-resonant Thomson scattering and resonant scattering. The latter is traditionally called `anomalous' scattering:
\begin{equation} \label{eq:fc}
f_\mathrm{c}=-r_e\left( (\boldsymbol{\epsilon'^{*}}\cdot\boldsymbol{\epsilon})F_\mathrm{T}(\mathbf{Q})+\boldsymbol{\epsilon'^{*}}\cdot\mathbf{F}_\mathrm{res}(\mathbf{Q})\cdot\boldsymbol{\epsilon}\right)
\end{equation}
where scalar $F_\mathrm{T}(\mathbf{Q})$ is the structure factor of the electron density and tensor $\mathbf{F}_\mathrm{res}(\mathbf{Q})$ is the structure factor of resonant scattering. 
The tensorial properties of $\mathbf{F}_\mathrm{res}(\mathbf{Q})$ are especially pronounced near absorption edges \cite{SIFinkelstein} where the resonant atomic factors 
reflect the symmetry of corresponding atomic positions, the shape and orientation of chemical bonds {\it etc}. For forbidden reflections $F_\mathrm{T}(\mathbf{Q})=0$ because scalar scattering 
of two (or more) equivalent atoms in the unit cell cancels exactly, whereas 
$\mathbf{F}_\mathrm{res}(\mathbf{Q})$ can be non-zero  because the tensor atomic factors of those equivalent atoms can be different if the atoms 
are related by screw-axes or glide mirror planes, {\it i.e.} they are oriented differently.

\begin{figure}[!h] 
\begin{center}
\includegraphics[angle=0,width=80mm]{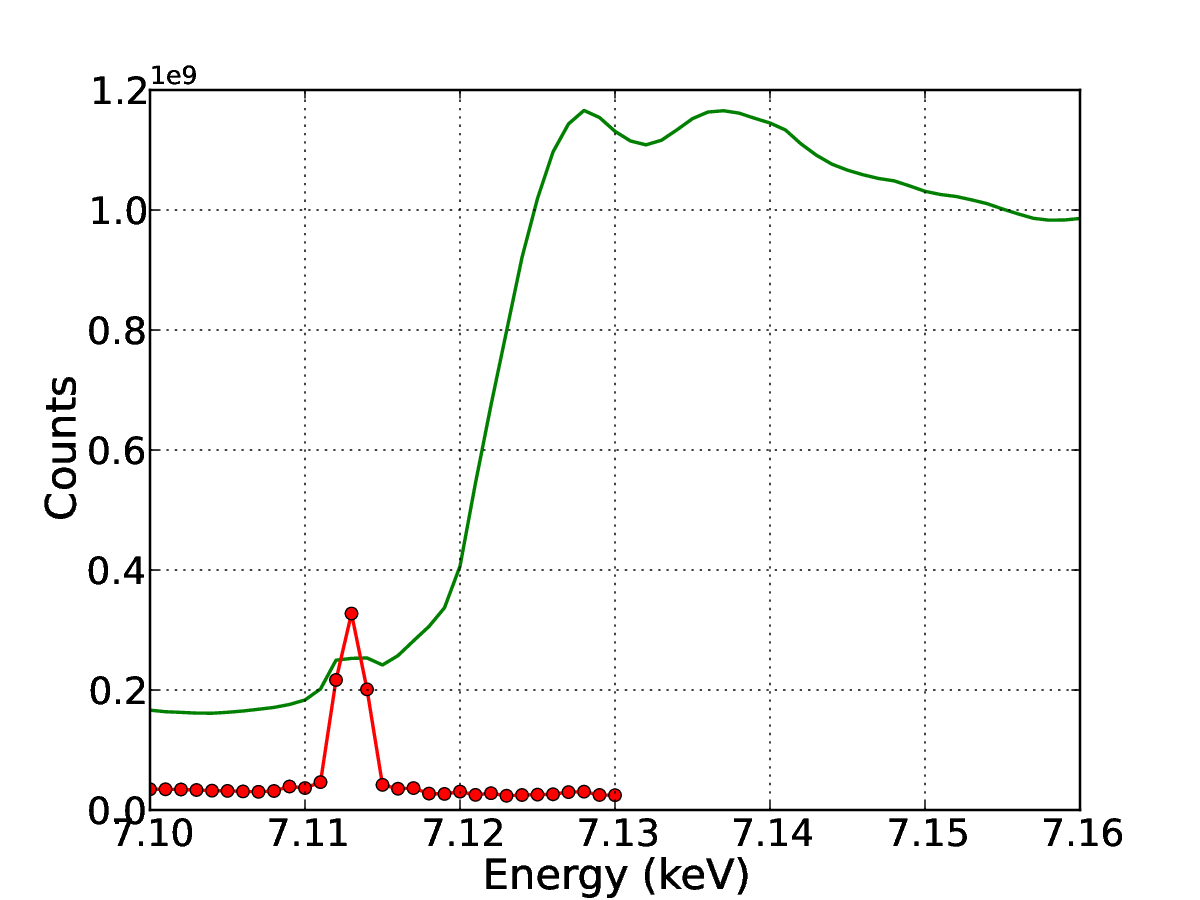}
\caption{\small
The absorption (fluorescence) spectrum of FeBO$_3$ near the Fe $K$ edge (green line), showing a weak pre-edge peak. Red: the
resonant (003) forbidden scattering spectrum, showing a single sharp peak at the absorption pre-edge position.
}
\label{FeBO3_energy_scan_I16}
\end{center}
\end{figure}

A symmetry analysis of the resonant scattering tensor \cite{SIDmitrienko1,SILovesey1} from the iron sites in FeBO$_3$ reveals that, for forbidden reflections of the form
$(hkl)=(0,0,6n+3)$, the lowest order contributing tensor is of rank four, which can arise from pure electric quadrupole $1s\rightarrow3d$ transitions. Since
transitions of this kind, into the relatively narrow 3$d$ band, tend to be reduced in energy due to differences in core-hole screening, one might expect a very sharp resonance
just below the Fe $K$ edge. This is precisely what was observed and is shown in Fig.~\ref{FeBO3_energy_scan_I16}, with experimental data taken again at Diamond I16.
That the tensorial (geometrical) properties of the resonant scattering are as expected can be confirmed by measuring the (003) resonant Bragg peak as a function of the 
sample azimuthal ($\psi$) rotation, and comparing with the calculated angle dependence. While multiple scattering events inevitable appear as `noise' in such a measurement,
the results indicate very good agreement with the calculations and give confidence in the physical interpretation of the scattering process.

The resonant scattering energy spectrum shows, to a first approximation, a single resonance. However, modelling the scattered intensity with the FDMNES program \cite{SIJoly01}
reveals an energy dependence that is slightly better described by two nearby resonances. The FDMNES results, verified by the experimental data, provide a valuable
tool to describe the resonant scattering.

\begin{figure}[!h] 
\begin{center}
\includegraphics[angle=0,width=100mm]{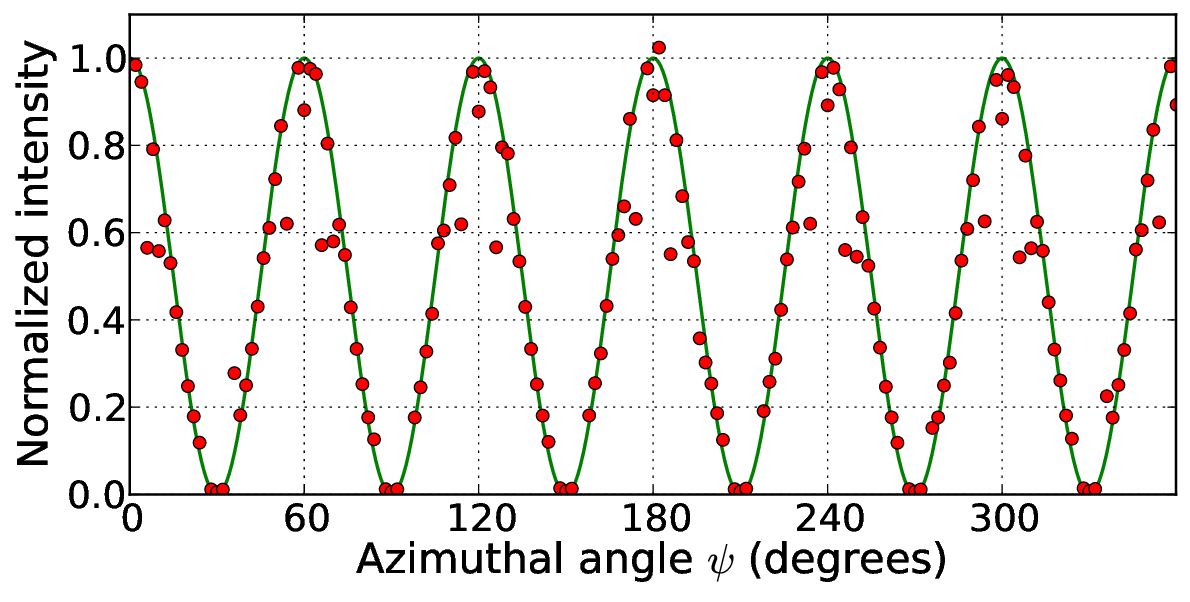}
\caption{\small
The measured sample azimuthal angle dependence of the (003) resonant scattering peak (red circles) along with a calculated
curve, assuming a forth-rank scattering tensor arising from pure electric quadrupole transitions.
}
\label{FeBO3_psi_scan_I16}
\end{center}
\end{figure}

It is interesting to note that, while weak ferromagnetism and resonant forbidden diffraction are each relatively rare, there
appears to be a very strong tendency for the former to exhibit the latter. This can be understood qualitatively by the fact they 
both phenomena rely on a twisted local environment of heavy atoms.

\section{Theory: FDMNES and scattering phases}
\begin{figure}[!h] 
\begin{center}
\includegraphics[angle=0,width=80mm]{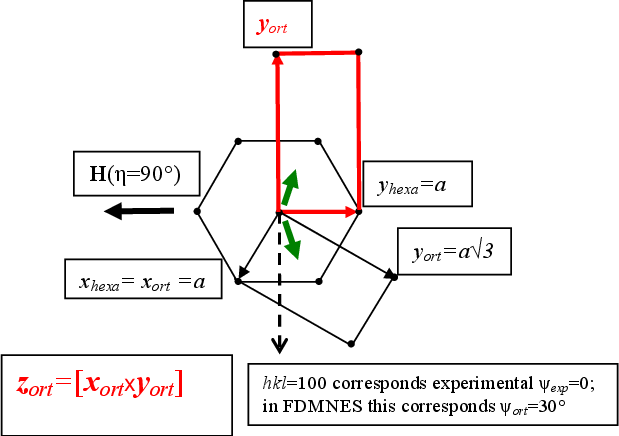}
\caption{\small
 Angles in FDMNES and in the experiment. Small circles are Fe atoms.
}
\label{angles_theory_expt}
\end{center}
\end{figure}

\begin{figure}[!h] 
\begin{center}
\includegraphics[angle=0,width=160mm]{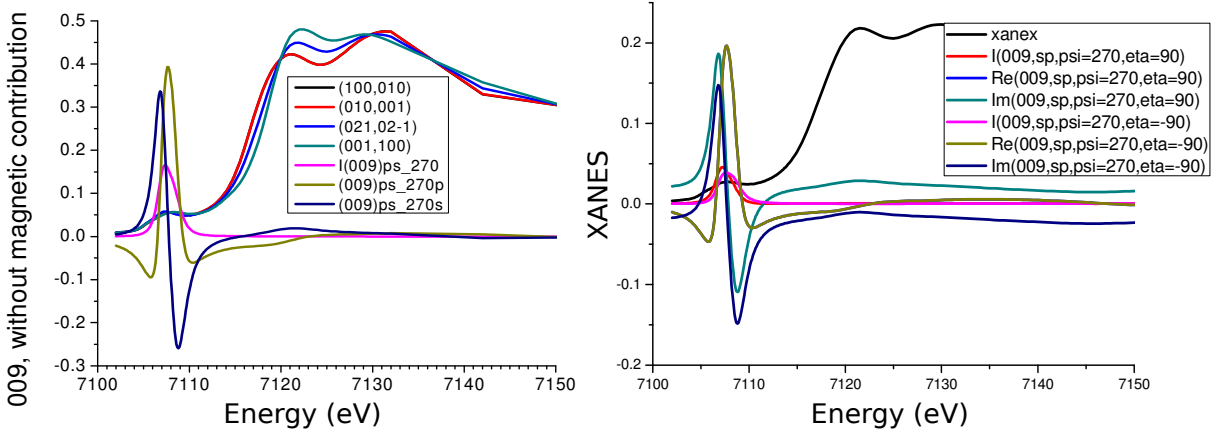}
\caption{\small
FDMNES simulations. Left: Real (green) and imaginary (dark blue) scattering amplitudes, and intensity (magenta), of (009) reflection for $\psi=270^{\circ}$
($\psi_{expt}=0$), along calculated absorption spectra for various crystallographic directions.
Right: The same intensities of (009) reflection for two orientations of magnetic field, shown with their real and imaginary parts.
The non-resonant magnetic scattering amplitude is constant, shifting the imaginary parts up or down. The real parts are almost indistinguishable.
}
\label{FDMNES_1}
\end{center}
\end{figure}

\begin{figure}[!h] 
\begin{center}
\includegraphics[angle=0,width=80mm]{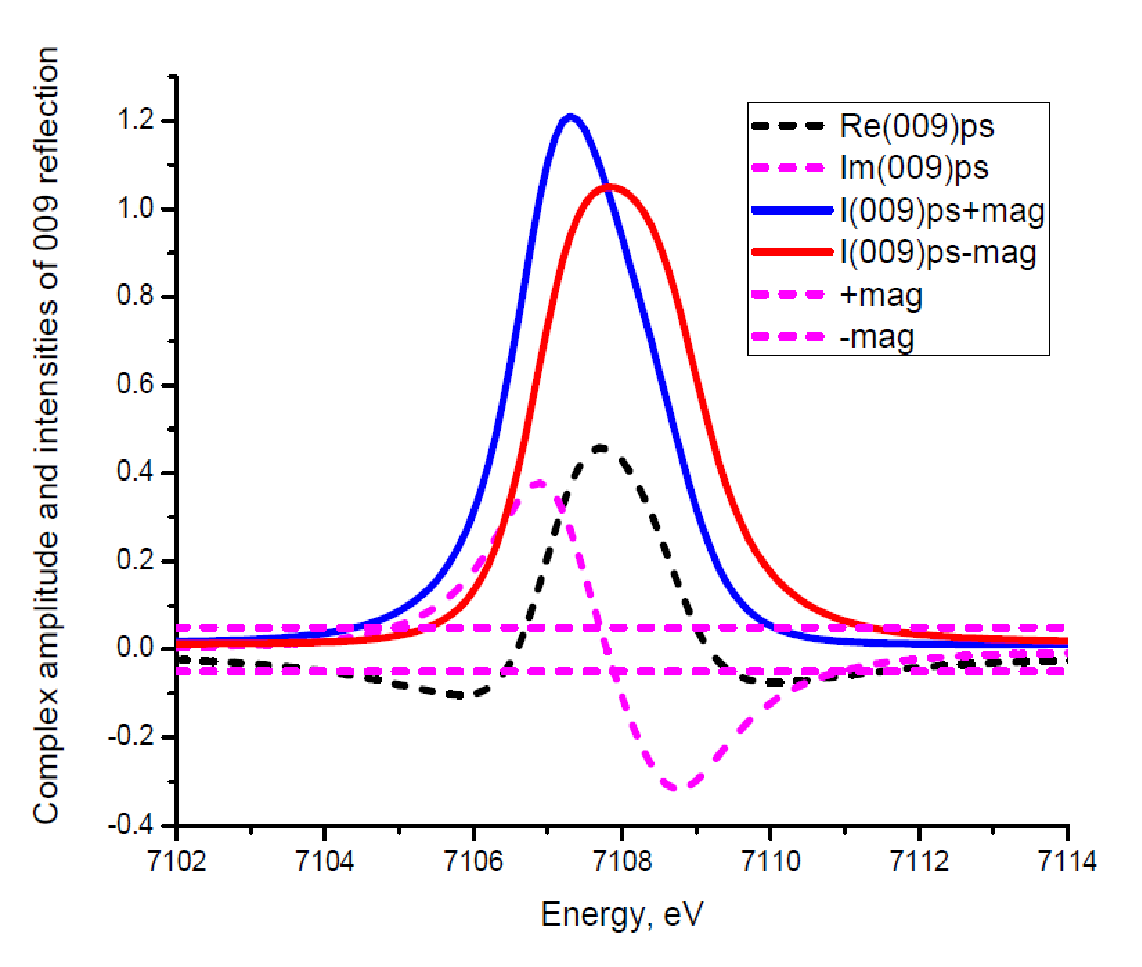}
\caption{Complex  scattering amplitudes and intensities of  the $\boldsymbol{\sigma}$-to-$\boldsymbol{\pi}$ (009) forbidden reflection, calculated using 
the FDMNES program. 
The black and magenta dotted lines correspond to the real and imaginary parts of the resonant structure factor in the pre-edge 
region just below the Fe $K$-edge. Straight magenta lines show pure imaginary amplitudes of non resonant (energy independent) 
magnetic scattering amplitudes which are positive for $\eta$=0 and negative for $\eta$=-60$^\circ$. The blue and red lines show intensities 
(square modulus of amplitudes) for $\eta$=0 and  $\eta$=-60$^\circ$ respectively.}
\label{FDMNES_2}
\end{center}
\end{figure}

Most crucial for the present study is the relative phase between the resonant charge and magnetic scattering terms, determined
by the structure factors $\mathbf{F}_\mathrm{res}(\mathbf{Q})$ and $\mathbf{F}_\mathrm{spin}(\mathbf{Q})$. 
The magnetic term was calculated according to Eq. \ref{eq:fm} and for resonant scattering amplitude we used the FDMNES program
(see Fig. \ref{angles_theory_expt} for some important definitions)
as outlined above. 
Fig.~\ref{FDMNES_1} shows the calculated absorption and scattering amplitudes with various azimuthal and field angles, while in 
Fig.~\ref{FDMNES_2} we highlight the real and imaginary parts of the resonant structure factor in the 
pre-edge region just below the Fe $K$-edge (black and magenta dotted lines correspondingly) and show how the intensity peak shifts down or up in energy due to 
interference with the positive or negative magnetic scattering amplitude. 
We see that a modest non-resonant 
magnetic scattering produces a pronounced difference in reflection intensities owing its interference with the imaginary
part of the resonant scattering. The most important aspect of the calculated magnetic/resonance interference is that we 
observe a clear energy jump which determines the phase of the magnetic scattering elegantly and unambiguously.

Let us assume (as predicted by the present theory) that the DM interaction induces a small left-hand twist of opposing spins of atoms at (0,0,0)
and (1/2,1/2,1/2). This means that in the DM energy, $E_{DM}={\mathbf D}\cdot[\mathbf{S}_0 \times \mathbf{S}_{1/2}]$, 
vector $\mathbf{D}$ is directed 
opposite to $\mathbf{z}$ (opposite to $\mathbf{c}$).
(Note that the magnetic moment is opposite to the spin direction since $g\simeq-2$ is negative). In Fig. \ref{angles_theory_expt} , for the indicated direction 
of $\mathbf{H}(\eta=90^{\circ})$,
the magnetic moment of the atom at (0,0,0) is directed up and slightly left whereas for the atom at (1/2,1/2,1/2) the magnetic moment 
is directed down and slightly left. Correspondingly, the spin of the atom at (0,0,0) is directed down and slightly right whereas for 
the atom at (1/2,1/2,1/2) its spin is directed up and slightly right (shown by short green arrows). For the opposite direction of $\mathbf{H}$ all 
the moments and spins change sign. The conventional orthorhombic unit cell (used in FDMNES) is shown as black rectangle. 
Experimental azimuthal angle $\psi_{exp}$=0 corresponds, in FDMNES,  to  $\psi_{ort}=30^{\circ}$ (dashed arrow); for  $\psi_{exp}=-60^{\circ}$, 
$\psi_{ort}=-30^{\circ}$.

It is convenient to rotate the orthogonal axes to an equivalent orientation so that $\mathbf{y}_{ort}$ will be vertical (red rectangle).
In this case the spin of the atom at (0,0,0) is directed along -$\mathbf{y}_{ort}$ whereas for the atom at (1/2,1/2,1/2) it is directed along $\mathbf{y}_{ort}$,
so that the spin structure factor $\mathbf{F}_{spin}$ is proportional to  -$\mathbf{y}_{ort}$. Now, for $\psi_{exp}=0$ the FDMNES azimuthal angle 
$\psi_{ort}=270^{\circ}$ 
or $-90^{\circ}$.

The FDMNES amplitudes are calculated for small cluster of radius 4.0 \AA, containing 33 atoms. We find that the 
size of the cluster and the method
of calculation (Green's function or finite difference method) are not crucial for the important details of quadrupole-quadrupole scattering, either in amplitude or
sign.
The sign of the non-resonant magnetic scattering in FDMNES is given by 
$f_{mag}=i\mathbf{B}\cdot \mathbf{F}_{spin}$, 
where {\bf B} is the vector of the non-resonant spin scattering. 
$\mathbf{B_{\pi^{\prime}\sigma}}=2\mathbf{k}\sin^2 \theta$, giving $f_{mag}=i|\mathbf{S}|\sin \eta$. 
For Thomson scattering FDMNES gives correctly 
$f_0+f^{\prime}+if^{\prime \prime}$ with $f_0+f^{\prime}<0$ and $if^{\prime \prime}>0$. 
(Note that, the magnetic scattering intensity in Fig. \ref{FeBO3_vs_fieldangle_new} does not go exactly to zero because the data correspond to the 
total scattering intensity $I_{\sigma^{\prime}\sigma}+I_{\pi^{\prime}\sigma}$, whereas we consider here only the rotated ($\pi^{\prime}\sigma$) polarization channel that
takes part in interference.)

\section{Theory: Electronic Structure Calculations}

To simulate the electronic structure and magnetic properties of iron borate in the ground state we used the tight-binding linear-muffin-tin-orbital 
atomic sphere approximation (TB-LMTO-ASA) method \cite{SItb-lmto} in terms of local spin density approximation, taking into account Hubbard U (LSDA+U) 
\cite{SIldau-Anisimov91} and spin-orbit coupling  (LDA+U+SO)  \cite{SIshorikov-lda-u-so}. 
The crystal structure data were taken from the literature \cite{SIdiehl}.
The radii of atomic spheres have been set to r(Fe)= 1.45 \AA, r(B)= 0.74 \AA ~ and r(O)= 0.85 \AA. In order to fill the interstitial 
space in the unit cell the required number of empty spheres was added.  

\begin{figure}[!h]
\begin{center}
\includegraphics[angle=270,width=100mm]{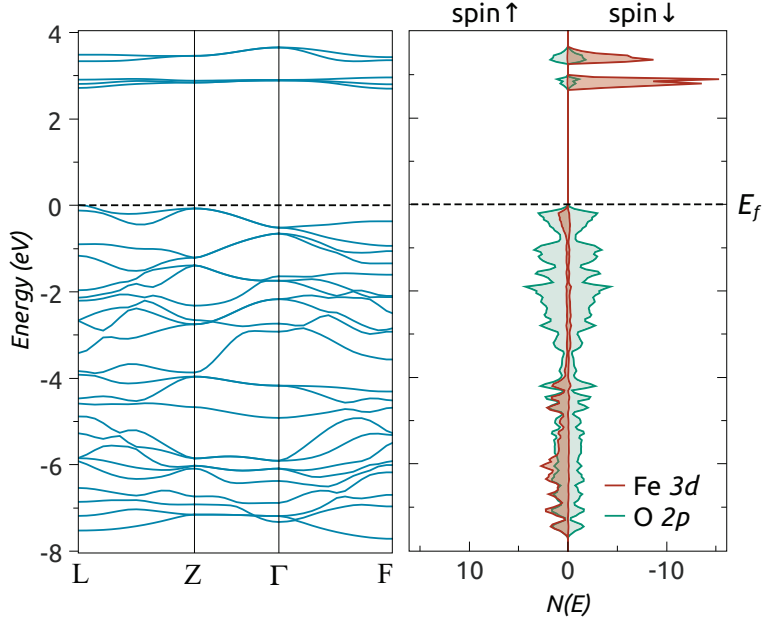}
\caption{Band structure (left panel) and partial densities of states (right panel) in iron borate obtained from LSDA+U calculation with U= 5 eV. 
Partial DOSs of Fe 3$d$ and O 2$p$ are shown by red and green colours correspondingly. DOSs originating from iron atoms with antiparallel 
magnetic moments (not shown here) are mirrored replicas of Fe 3$d$ DOS with swapped spin channels.}
\label{FigDOS}
\end{center}
\end{figure}

Previous theoretical investigations aiming at the description of the insulating state of iron borate demonstrated that the choice of the on-site 
Coulomb interaction, U plays an important role in reproducing the correct value of the bandgap of 2.8 eV \cite{SIfebo3-band-gap1,SIfebo3-band-gap2}. 
Depending on the method the value of the U parameter varies from 2.97 eV (many-body model calculations) to 7 eV (first-principles calculations) 
\cite{SIfebo3-ovchinnikov,SIfebo3-ggau}. In the present work we focus on the correct description of the magnetic couplings between iron moments and 
the value of the canting angle observed in experiment. As we will show below, good agreement can be achieved with U=5 eV.  
The same value was used to reproduce the electronic and magnetic structure of haematite ($\alpha$-Fe$_{2}$O$_{3}$) -- another classical 
example of antiferromagnet with weak ferromagnetism \cite{SImazurenko-fe2o3}.

Fig. \ref{FigDOS} gives the band structure and partial densities of states (DOSs) calculated using the LSDA+U approach for antiferromagnetic (AFM) 
configuration (without spin canting). The electronic spectrum agrees with that presented in previous works \cite{SIfebo3-ggau}. One can see 
that the unoccupied bands above the Fermi level demonstrate localized behaviour. They mainly originate from the iron 3$d$ states. This is 
not the case for valence band where a strong hybridization of the 3$d$ iron and 2$p$ oxygen states is observed. The LSDA+U value of the 
energy gap of 2.7 eV is in excellent agreement  with experimental value.

The previous ab-initio investigations \cite{SIfebo3-ggau} revealed two possible magnetic solutions with high-spin and low-spin ground states.  
In our calculations we can reproduce both of them by varying the value of the initial spin splitting for the on-site LSDA+U potential. 
The resulting magnetic moments are $M_{Fe}$ = 4.28 $\mu_{B}$ for high-spin ($S=\frac{5}{2}$) and $M_{Fe}$ = 1.1 $\mu_{B}$ for low-spin 
$S=\frac{1}{2}$) states. Below we will consider the high-spin solution which is the ground state at ambient conditions (and corresponds 
to experimental data).  The calculations performed for the ferromagnetic configuration of iron moments demonstrate a finite magnetization 
of the oxygen atoms ($\sim$ 0.2 $\mu_{B}$), which is another manifestation of  a strong hybridization between iron and oxygen states. 

\section{Theory: Isotropic Magnetic Coupling}

Having described the antiferromagnetic ground state without spin-orbit coupling we are going to analyse the low-energy magnetic excitation 
spectrum of the system to obtain information about isotropic exchange (IE) interactions for the Heisenberg Hamiltonian, {\it i.e.} the first term in Eq.~(1)(main text). 
For this purpose, one can use a magnetic force theorem that, being  formulated in terms of Green's functions, have produced an array of 
very reliable results concerning magnetic couplings in transition metal oxides \cite{SIsolovyev-mno,SImazurenko-na2v3o7}. Starting from the 
collinear ground state the exchange interaction parameters are determined via calculation of second variation of total energy  
$\delta^2 E$ for small deviation of magnetic moments, 
$J_{mn} = \delta^2 E / \delta \mathbf{S}_{m} \delta \mathbf{S}_{n}$.
The advantage of this method is that the expression for the second derivative can be derived analytically and requires for its 
evaluation only calculation of the integral over the product of the one-electron Green's functions.

The isotropic exchange interactions between magnetic moments calculated for U= 5 eV are presented in Table \ref{TabExch}.  
It shows that there is a strong interaction of the central site $0$ and iron atoms that belong to the first coordination sphere.
The couplings with next nearest neighbours are at least of one order of magnitude smaller and we can neglect them.  The leading 
magnetic interaction of 10.3 meV is in good agreement with experimental estimate of 7.5 meV in \cite{SImoessbauer-febo3}.
These results, obtained with the magnetic force theorem, can be confirmed by the total energy difference method where the exchange 
interaction defined \cite{SIexch-via-diff} as $J = (E_{FM} -E_{AFM}) / 4 z $. Here $E_{FM}$ ($E_{AFM}$) is energy of 
the ferromagnetic (antiferromagnetic) configuration and $z=6$ is a number of nearest neighbours.  
We obtain $J$ =9 meV, which is in good agreement with the Green's function method results. 

\begin{table}[!h]
\centering
\caption [Bset]{Calculated values of isotropic exchange interactions between magnetic moments in FeBO$_3$ (in meV). 
The number in parentheses denotes the coordination sphere. }
\bigskip 
\label{TabExch}
\begin {tabular}{lcccccc}
 \hline
 Fe$^{(1)}$ & Fe$^{(2)}$ & Fe$^{(3)}$ & Fe$^{(4)}$       & Fe$^{(5)}$            & Fe$^{(6)}$  & Fe$^{(7)}$   \\
  10.28 & 0.21 & 0 &0.54 & -0.08 & 0 & 0.02 \\
 \hline
\end {tabular}
\end {table}

\section{Theory: Modeling the canted state}
The experimental investigations revealed that in the ground state there is a canting of the magnetic moments with respect to the antiferromagnetic configuration. 
The value of the canting of
0.96$^{\circ}$
is one order of magnitude larger than that observed in other antiferromagnets with weak ferromagnetism such as 
Fe$_{2}$O$_{3}$ and La$_{2}$CuO$_{4}$ \cite{SImazurenko-fe2o3}. To describe this non-collinear magnetic state we performed LDA+U+SO \cite{SIshorikov-lda-u-so} 
calculations that incorporate spin-orbit coupling. The latter leads to an orbital magnetism and is responsible for magnetocrystalline single-site
anisotropy and Dzyaloshinskii-Moriya interactions between magnetic moments. Depending on initial directions of the magnetic moments in our calculations we 
simulated different magnetic configurations. Two stable magnetic states, with moments lying along $x$ and $z$ axes, were obtained. The main difference between 
these solutions is that we observe a canting of the spins when moments are lying in the 
$xy$ plane (Table \ref{TabSpin}). This is not the case for $z$-oriented configurations where they are strictly antiparallel. The LDA+U+SO value of the canting 
angle,  $\phi_{LDA+U+SO}$ = 0.96$^{\circ}$
 is in excellent  agreement with the experimental estimate. The energy difference between $x$- and $z$-type magnetic configurations 
 $E^{x}_{total} - E^{z}_{total} = - 0.07$ meV indicates that the non-collinear canted configuration corresponds to the minimum of the system energy. 
\begin {table}[!h]
\centering
\caption [Bset]{Absolute values ($\mu_{B}$)  and directions of spin and orbital magnetic moments of iron atoms in the unit cell. These results 
were obtained from the self-consistent LDA+U+SO calculations. } 
\bigskip
\label{TabSpin}
\begin {tabular}{ccccc}
  \hline
Fe($i$) & moment & e$_x$  & e$_y$ & e$_z$ \\
  \hline
A & $|\vec S|$ = 4.249 & -0.999 & -0.016 & 0 \\
  & $|\vec L|$ = 0.027 & -0.999 &  -0.017 &  0 \\
  \hline
B & $|\vec S|$ = 4.249 & 0.999 & -0.016 & 0 \\
  & $|\vec L|$ = 0.027 & 0.999 &  -0.017 &  0 \\
  \hline
\end {tabular}
\end {table}

Using the calculated magnetic moments (Table II) we can obtain the sign and value of some components of the Dzyaloshinskii-Moriya interaction in the 
Hamiltonian given by Eq. 1 (main text).
Since the vector product $[\mathbf{S}_{1} \times \mathbf{S}_{2}]$ is parallel to z-axis, the corresponding anisotropic coupling must be antiparallel in order to
minimize the energy of the system, {\it i.e.}, $D^{z}_{AB} < 0$. As for the absolute value of the DM interaction, it can be estimated via canting angle and isotropic exchange 
interaction, $|D^{z}_{AB}| = 2 J| S^{y}/S^{x}| $ =   0.33 meV. Here $J$ = 10.3 meV, the isotropic exchange interaction of the 
Fe atom with the nearest neighbours. Thus the estimated {\it z}-component of the DM interaction can be associated with the individual 
antisymmetric
exchange interaction 
of the 0$^{th}$ site with atoms belonging the first coordination sphere.

\section{Theory: Correlated Band Method for calculating the Dzyaloshinskii-Moriya parameters}

Here, we present a second, and very general, method for the calculation of individual DM interactions.
There are two important methodological steps in our approach. First, we consider the effect of the local rotations $\delta \boldsymbol{\varphi}_{m}$ 
of the total 
angular momentum
operator, 
\newline
$\hat R_{m} = \exp(i\delta \boldsymbol{\varphi}_{m}\cdot \hat{\mathbf{ J}}_{m})$, on the inter-site (hopping) part 
of the Hubbard Hamiltonian with rotationally invariant form of the Coulomb interaction, $\hat H'_{mn} = \hat R^{+}_{m}  \hat H_{mn} \hat R_{n}$. 
The second step is to integrate out the fermionic degrees of freedom in the expression for the variation of the total energy.
This is an adiabatic approximation \cite{SIantropov} where we assume that  the spin dynamics with the typical energy scale varying from 0.05 meV 
(DM interaction) to 10 meV (isotropic exchange interaction) is much slower than the electronic processes with characteristic energies of intra-atomic exchange or bandwidth ($\sim$1  eV). 
The actual time scales of spin and electronic processes in a particular strongly correlated system are accessible with advanced experimental techniques. 
In this respect an important information about spin dynamics in FeBO$_{3}$ due to the ultrafast laser impulse was 
reported in the literature \cite{SIRasing1, SIRasing2}.

\begin{figure}[!h] 
\begin{center}
\includegraphics[angle=0,width=110mm]{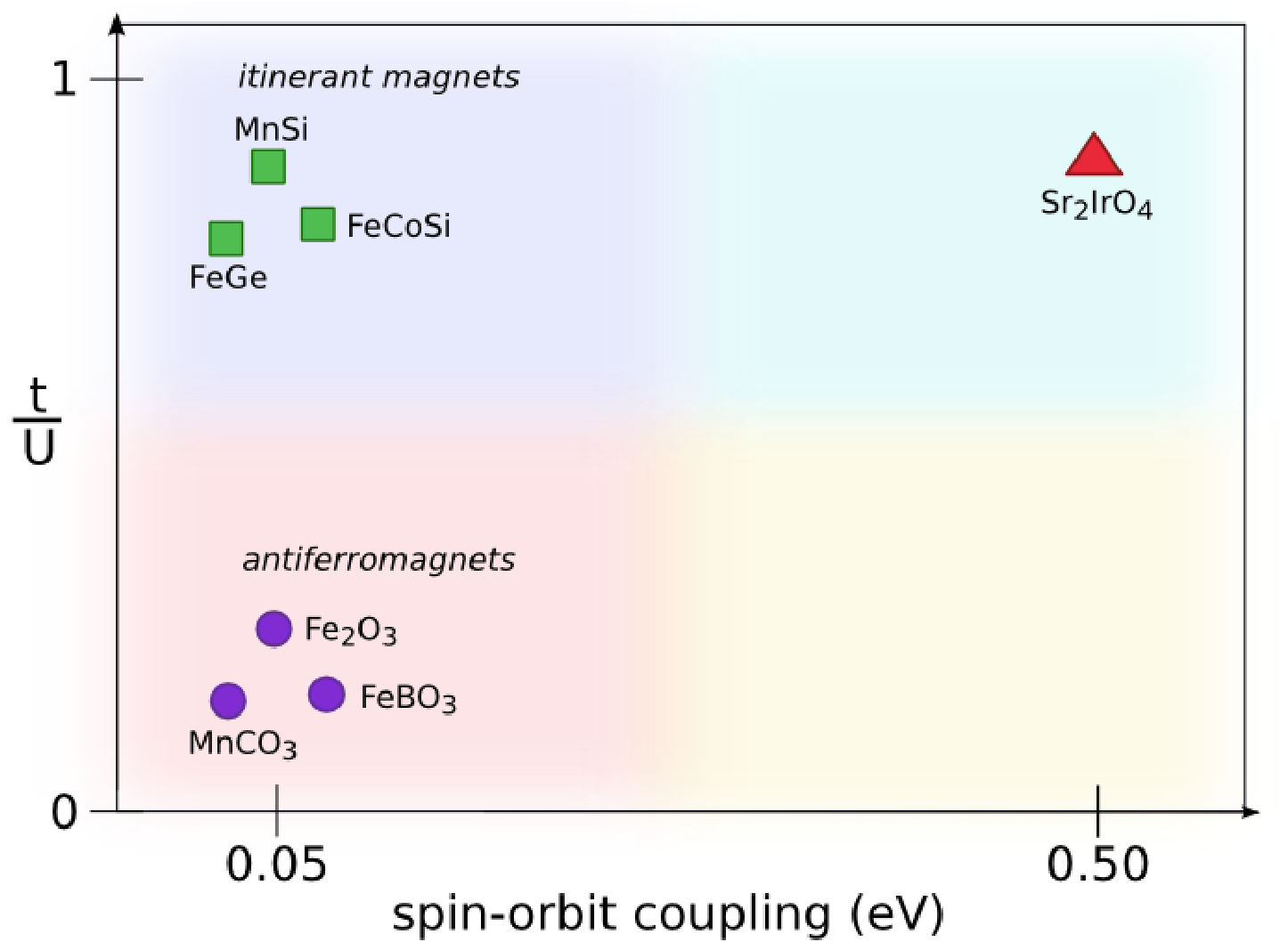}
\caption{Classification of Dzyaloshinskii-Moriya materials with respect to the strength of correlation and spin-orbit coupling effects. 
The ratio between the band width ($t$) and on-site repulsion (U) shows the degree of correlations.
The coloured regions denote working space for correlated band DM method.}
\label{Phasediag}
\end{center}
\end{figure}

The resulting variation of the electronic Hamiltonian has a very compact form and contains the Dzyaloshinskii-Moriya interaction (anticommutator) and symmetric anisotropic exchange interaction (commutator), 
\begin{eqnarray}
\delta E = Ê-\frac{i}{2} \sum_{mn}(\delta \vec \varphi_{m} -
\delta \vec \varphi_{n}) Tr_{L, \sigma} \biggl\{N_{nm} [\hat{\mathbf{J}}, \hat t_{mn}]_{+} \biggl\}  \nonumber \\
-\frac{i}{2} \sum_{mn}(\delta \vec \varphi_{m} +
\delta \vec \varphi_{n}) Tr_{L, \sigma} \biggl\{N_{nm} [\hat{\mathbf{J}}, \hat t_{mn}]_{-} \biggl\},
\label{rotE}
\end{eqnarray}
where $N_{nm}$ is the energy-integrated inter-site Green's function (occupation matrix) which describes the propagation of an electron from site $n$ to $m$, 
$\hat{\mathbf{ J}}$ is the total moment operator and $t_{mn}$ is the hopping  matrix. 
Being formulated in the Wannier function basis this approach naturally takes into account the hybridization between metal and ligand 
atoms of the strongly correlated system - important for simulation of itinerant magnets where there is a strong  delocalization 
of the magnetic moment. 
All matrices are in spinorial form and corresponding 
spin and orbital indices are omitted. From Eq.~(\ref{rotE}) one clearly sees that DM interaction  is an antisymmetric interaction of kinetic origin. 
The use of the total angular momentum operator allows us to probe the spin and orbital structure of the hopping matrix. 

In contrast to other numerical approaches \cite{SImazurenko-fe2o3, SIsolovyev-lamno3, SIrudenko} for calculating DM interaction which are based on the 
rotation of the on-site exchange fields, our method has fully inter-site formulation (i.e. the rotation of the hopping matrix). 
For the case of orbitally-independent magnetic excitations the on-site and inter-site considerations provide two alternative ways to solve the problem of the
determination of spin Hamiltonian parameters. 
On the other hand our inter-site approach becomes preferable when the potential depends on the particular $3d$ orbital, and spin-orbit coupling is 
taken into account. In this case the `on-site' formulation leads to a complicated situation where, in general, the spins of different orbitals can 
be noncollinear to each other (intra-atomic noncollinear magnetic ordering) \cite{SIantropov}, requiring 
the spin dynamics for different orbital states to be treated independently  at one site. In turn, such an orbital resolution complicates the resulting 
magnetic model that will also have orbital degree of freedom. As we previously demonstrated \cite{SIkatsnelson} our inter-site consideration allows 
one to preserve the simplicity of the magnetic model while taking into account orbital and spin excitations of the system. 
Thus, the correlated band method for calculating DM interaction parameters can be used for a wide range of materials (Fig. \ref{Phasediag}) with different strengths of the spin-orbit coupling and correlation effects.

For FeBO$_{3}$,  the calculated  anisotropic exchange interactions between atom $0$ and atoms that belong to the first coordination sphere 
(Fig. 1, main text) are presented in Table \ref{TabDM}. They are not independent and can be transformed to each other by using the symmetry operations of the space group $R\bar{3}c$. For instance, the transformation from the bond 0-1 to 0-2 corresponds to the rotation on 120$^{\circ}$ around x-axis. Such an operation 
changes neither
the {\it z}-component of the position vector nor the Dzyaloshinskii-Moriya vector.  There is also an inversion centre at the 0$^{th}$ site, which means that the following relations  are satisfied, $\vec D_{01} = \vec D_{04}$, $\vec D_{02} = \vec D_{06}$ and $\vec D_{03} = \vec D_{05}$. From Table \ref{TabDM} one can see that our numerical method reproduces these relations with a good accuracy and that the sum of all DM vectors lies along the three-fold axis, $z$. Moreover, the sign of DMI fully corresponds to the canted ground state of FeBO$_{3}$ obtained in LDA+U+SO calculations. 
\begin {table}[!h]
\centering
\caption [Bset]{Parameters of Dzyaloshinskii-Moriya interaction (in meV) calculated by using Eq.(\ref{rotE}).} 
\label{TabDM}
\bigskip 
\begin {tabular}{ccc}
  \hline
Bond $m-n$ & $\mathbf{R}_{mn}$ & $\mathbf{D}_{mn}$ (meV) \\
\hline
0-1 & (1.0 ; 0.0 ; -0.9044) 		&  (-0.249; 0.0; -0.240) \\
\hline
0-2 & (-0.5 ; -$\sqrt{3}/2$ ; -0.9044)  & (0.124 ; 0.216 ;  -0.240) \\
\hline
0-3 & (-0.5 ; $\sqrt{3}/2$ ; -0.9044)   &   (0.124 ;  -0.216 ;  -0.240) \\
\hline
0-4 & (-1.0 ; 0.0 ; 0.9044) 		&  (-0.249; 0.0 ; -0.240) \\
\hline
0-5 & (0.5 ; -$\sqrt{3}/2$ ; 0.9044)    &  (0.124 ;  -0.216 ;  -0.240) \\
\hline
0-6 & (0.5 ; $\sqrt{3}/2$ ; 0.9044)     & (0.124 ; 0.216 ;  -0.240) \\
\hline
\end {tabular}
\end {table}

Having calculated all the parameters of the spin Hamiltonian we are in position to describe the weak ferromagnetism observed in iron borate.
For this purpose, one should define the canting angle and the plane of the spin rotation, $\delta \boldsymbol{\phi}$, which can be done in the mean-field approximation,
\begin{eqnarray}
\delta \boldsymbol{\phi} = \frac{\sum_{n} \mathbf{D}_{0n}}{2 \sum_{n} J_{0n}}
\end{eqnarray} 
Here the summation runs over the atoms that belong to the first coordination sphere.
The  symmetry of the canting is defined by the symmetry of the magnetic torque, i.e. the summary DMI. 
In our case the latter is along $z$ axis, which means the rotation of the spins occurs in the $xy$ plane. The value of $|\boldsymbol{\phi} |$ calculated this way is equal to 0.7$^{\circ}$, which is only slightly smaller than the value obtained above by  self-consistent calculation and the experimental value of 0.9$^{\circ}$.

\end{document}